\documentclass[runningheads]{llncs}
\usepackage[T1]{fontenc}

\usepackage{bold-extra}
\usepackage{thmtools} 
\usepackage{thm-restate}
\usepackage{lipsum}

\newcommand{\OPT}{\operatorname{OPT}}
\newcommand{\DP}{\operatorname{DP}}

\newenvironment{claimproof}
  {%
   \begin{proof}}
  {\end{proof}}

\usepackage{tikz}
\usetikzlibrary{calc}

\usepackage[dvipsnames,table]{xcolor}
\usepackage{tcolorbox}
\tcbuselibrary{skins, breakable, theorems}

\usepackage[mathlines]{lineno}
\newcommand{\lineqed}{%
  \unskip\nobreak\hfill$\squareforqed$%
}

\usepackage{graphicx}
\usepackage{amsmath, amssymb, amsfonts}
\usepackage{todonotes}

\usepackage{tcolorbox}
\tcbuselibrary{skins}
\usepackage{xcolor}
\usepackage{mdframed}

\colorlet{mix}{red!50!black}

\usepackage{hyperref}
\hypersetup{colorlinks={true},linkcolor={mix},citecolor=blue}

\usepackage{cleveref} % Only for LNCS

\newcounter{sasanka}

\newlength{\RoundedBoxWidth}
\newsavebox{\GrayRoundedBox}
\newenvironment{GrayBox}[1]%
{\setlength{\RoundedBoxWidth}{.93\textwidth}
	\def\boxheading{#1}
	\begin{lrbox}{\GrayRoundedBox}
		\begin{minipage}{\RoundedBoxWidth}}%
		{   \end{minipage}
	\end{lrbox}
	\begin{center}
		\begin{tikzpicture}%
			\node(Text)[draw=black!20,fill=white,rounded corners,%
			inner sep=2ex,text width=\RoundedBoxWidth]%
			{\usebox{\GrayRoundedBox}};
			\coordinate(x) at (current bounding box. north west);
			\node [draw=white,rectangle,inner sep=3pt,anchor=north west,fill=white] 
			at ($(x)+(6pt,.75em)$) {\boxheading};
		\end{tikzpicture}
\end{center}}     

\newenvironment{defproblemx}[2][]{\noindent\ignorespaces%
	\FrameSep=6pt%
	\parindent=0pt%
	\vspace*{-1.5em}
	\ifthenelse{\isempty{#1}}{%
		\begin{GrayBox}{\textsc{#2}}%                
		}{%
			\begin{GrayBox}{\textsc{#2}  parameterized by~{#1}}%  
			}
			\begin{tabular*}{\textwidth}{@{\hspace{.1em}} >{\itshape} p{1.8cm} p{0.8\textwidth} @{}}%        
			}{
			\end{tabular*}%
		\end{GrayBox}%
		\ignorespacesafterend
	}

	\usepackage{xspace}
	\makeatletter
	\newcommand*{\rom}[1]{\expandafter\@slowromancap\romannumeral #1@}
	\makeatother

    \definecolor{segblue}{HTML}{0072B2}
\definecolor{segorange}{HTML}{E69F00}
\definecolor{seggreen}{HTML}{009E73}
\definecolor{segpurple}{HTML}{CC79A7}

	\newcommand{\yes}{\texttt{Yes}\xspace}

	\newcommand{\fpt} {{\sf FPT}}

    \newtheorem{theo}{Theorem}
	\newtheorem{clm}{Claim}

	\newtheorem{branching rule}{Branching Rule}
	\newtheorem{reduction rule}{Reduction Rule}
	\newtheorem{marking scheme}{Marking Scheme}

	\newtheorem{observation}[theorem]{Observation}
\newtheorem{reductionrule}{Reduction Rule}

\DeclareMathOperator{\relint}{relint}

\newcommand{\bigO}{\mathcal O}
\newcommand{\sco}{\zeta}
\newcommand{\lc}{\lambda}
\newcommand{\SC}{\textnormal{\textsc{Segment Cover}}}
\newcommand{\LRSC}{{{\sc Supported Segment Cover}}}
\newcommand{\aee}{\mathcal{A}}

\newcommand{\iee}{\mathcal{I}}

\newcommand{\lee}{\mathcal{L}}

\newcommand{\ree}{\mathcal{R}}
\newcommand{\see}{\mathcal{S}}

\definecolor{ccdBlue}    {HTML}{1B4F8A}   % deep navy  — header / frame
\definecolor{ccdLightBlu}{HTML}{D6E8F7}   % pale sky   — CCD box bg
\definecolor{ccdTeal}    {HTML}{0D7377}   % teal       — variant frames
\definecolor{ccdLightTea}{HTML}{D0EEEE}   % pale teal  — variant box bg
\definecolor{ccdGreen}   {HTML}{1A6B3C}   % forest green — tractable
\definecolor{ccdRed}     {HTML}{8B1A1A}   % deep red   — hardness
\definecolor{ccdGray}    {HTML}{F2F4F7}   % light gray — table stripe

\definecolor{ccdBlue}{HTML}{294C6B}
\definecolor{ccdLightBlue}{HTML}{F3F6F8}

\definecolor{ccdTeal}{HTML}{3F6F6A}
\definecolor{ccdLightTeal}{HTML}{F3F7F6}

\definecolor{ccdBurgundy}{HTML}{754B55}
\definecolor{ccdLightBurgundy}{HTML}{F8F4F5}

\tcbset{
  problembase/.style={
    enhanced,
    breakable,
    fonttitle=\bfseries\sffamily,
    coltitle=white,
    attach boxed title to top left={xshift=4mm,yshift=-2mm},
    boxed title style={
      arc=2pt,
      boxrule=0pt,
      left=5pt,
      right=5pt,
      top=2pt,
      bottom=2pt
    },
    arc=3pt,
    boxrule=0.8pt,
    left=6pt,
    right=6pt,
    top=9pt,
    bottom=6pt,
    before skip=10pt,
    after skip=10pt
  },
  masterbox/.style={
    problembase,
    colback=ccdLightBlue,
    colframe=ccdBlue,
    boxed title style={
      colback=ccdBlue,
      colframe=ccdBlue,
      arc=2pt,
      boxrule=0pt,
      left=5pt,
      right=5pt,
      top=2pt,
      bottom=2pt
    }
  },
  variantbox/.style={
    problembase,
    colback=ccdLightTeal,
    colframe=ccdTeal,
    boxed title style={
      colback=ccdTeal,
      colframe=ccdTeal,
      arc=2pt,
      boxrule=0pt,
      left=5pt,
      right=5pt,
      top=2pt,
      bottom=2pt
    }
  },
  relatedbox/.style={
    problembase,
    colback=ccdLightBurgundy,
    colframe=ccdBurgundy,
    boxed title style={
      colback=ccdBurgundy,
      colframe=ccdBurgundy,
      arc=2pt,
      boxrule=0pt,
      left=5pt,
      right=5pt,
      top=2pt,
      bottom=2pt
    }
  }
}

\usepackage{tabularx,lipsum,environ}
\usepackage{tikz}

\makeatletter
\newcommand{\problemtitle}[1]{\gdef\@problemtitle{#1}}% Store problem title
\newcommand{\probleminput}[1]{\gdef\@probleminput{#1}}% Store problem input
\newcommand{\problemquestion}[1]{\gdef\@problemquestion{#1}}% Store problem question
\NewEnviron{problemm}{
	\problemtitle{}\probleminput{}\problemquestion{}% Default input is empty
	\BODY% Parse input
	\par\addvspace{.5\baselineskip}
	\noindent
	\begin{tabularx}{\textwidth}{@{\hspace{\parindent}} l X c}
		\multicolumn{2}{@{\hspace{\parindent}}l}{\@problemtitle} \\% Title
		\textbf{Input:} & \@probleminput \\% Input
		\textbf{Question:} & \@problemquestion% Question
	\end{tabularx}
	\par\addvspace{.5\baselineskip}
}

\begin{document}
	
	\title{Faster Parameterized Algorithms for Covering Points with Non-Crossing Segments}
	\titlerunning{Faster Parameterized Algorithms for CPNCS}

%Covering Points with Non-Crossing Segments: Faster Parameterized  Algorithm
	\author{Satyabrata Jana\inst{1}, 
		 Debabrata Pal \inst{2}, Sasanka Roy \inst{2}}
	
	\authorrunning{S.~Jana, D.~Pal, S.~Roy}
	\institute{Indian Institute of Science Education and Research Berhampur, Brahmapur, India \and Indian Statistical Institute, Kolkata, India \\
		\email{\{satyamtma, debabratapal4521, sasanka.ro\}@gmail.com
	}}

%%
%\title{Contribution Title}
%%
%%\titlerunning{Abbreviated paper title}
%% If the paper title is too long for the running head, you can set
%% an abbreviated paper title here
%%
%\author{First Author\inst{1}\orcidID{0000-1111-2222-3333} \and
%Second Author\inst{2,3}\orcidID{1111-2222-3333-4444} \and
%Third Author\inst{3}\orcidID{2222--3333-4444-5555}}
%%
%\authorrunning{F. Author et al
%% First names are abbreviated in the running head.
%% If there are more than two authors, 'et al.' is used.
%%
%\institute{Princeton University, Princeton, NJ 08544, USA \and
%Springer Heidelberg, Tiergartenstr. 17, 69121 Heidelberg, Germany
\maketitle

\begin{abstract}

Two line segments are \emph{non-crossing} if their relative interiors are disjoint.  Given a finite set $P$ of $n$ points in the plane and a nonnegative integer $k$, the {\sc Covering Points with Non-Crossing Segments (CPNCS)} problem asks whether $P$ can be covered by at most $k$ pairwise non-crossing line segments.  Recently [Grelier, SoCG~2022], the problem was shown to be  {\sf NP}-hard and fixed-parameter tractable when parameterized by the solution size $k$, with running time $ \mathcal{O}(2^{k^2}k^{7k}+n^4\log n)$. A natural question is whether the dependence on $k$ can be improved asymptotically. We answer this in the affirmative by giving an {\sf FPT} algorithm for {\sc CPNCS} with running time $ k^{\bigO(k)}\,n^{\bigO(1)}. $

\par\smallskip
\hspace*{1em} We further study {\sc Supported CPNCS}, a variant in which the input also contains a prescribed family $\lee$ of $t$ lines, and every solution segment must be contained in some line of $\lee$.  We give an {\sf FPT} algorithm, parameterized by $t$, running in time $ 2^{\bigO(t)}n^{\bigO(1)} $ and obtain a  kernel containing at most
$\mathcal{O}(t^2)$ points.  We also show that when parameterized by $k$, {\sc Supported CPNCS} admits an {\sf FPT} algorithm  with running time $ k^{\bigO(k)}\,n^{\bigO(1)}. $

\par\smallskip
\hspace*{1em}Finally, we investigate the \emph{point line cover number} $\lambda(P)$, defined as the minimum number of lines whose union contains $P$.  Let $\sco(P)$ denote the minimum size of a non-crossing segment cover of $P$.  Since  $\lambda(P)\leq\sco(P)$, $\lambda(P)$ is a natural structural parameter for {\sc CPNCS}.  We show that {\sc CPNCS} admits an {\sf FPT} algorithm, when parameterized by $\lambda(P)$, with running time
$ \lambda(P)^{\bigO(\lambda(P)^2)}n^{\bigO(1)}$.

\end{abstract}

\keywords{Point-line cover  \and non-crossing segment  \and parameterized algorithms}

\section{Introduction}\label{sec:intro}

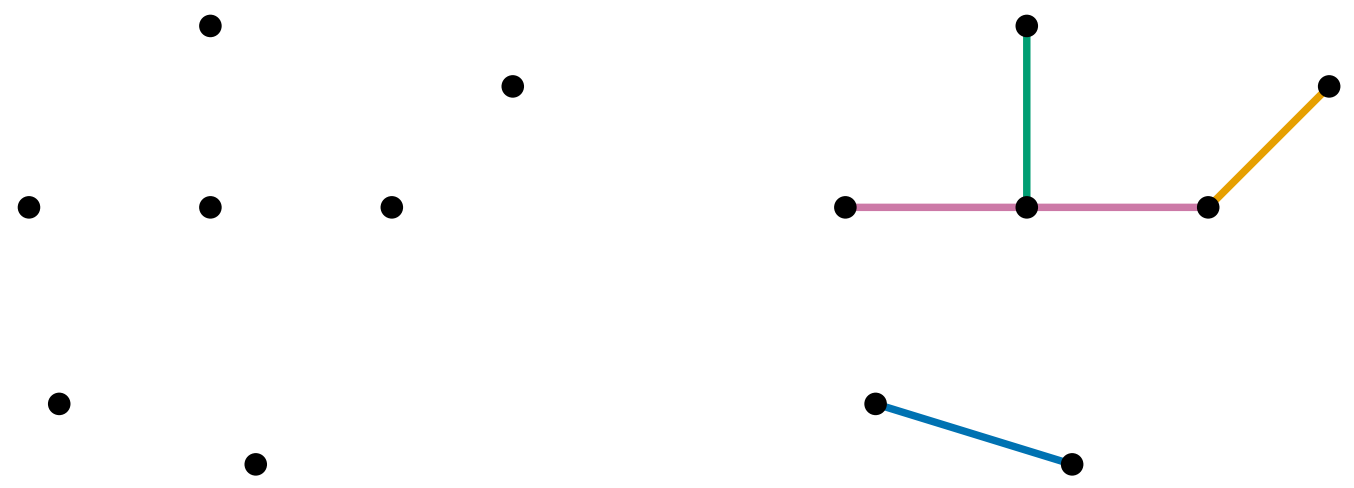
\begin{figure}[t]
\centering
\begin{tikzpicture}[
    dot/.style={circle,fill=black,inner sep=1.5pt},
    segment/.style={line width=1.4pt}
]

% =========================================================
% (a) Input point set
% =========================================================

\node[dot,label=below:] at (0,0.5) {};
\node[dot,label=below:] at (1.2,0.5) {};
\node[dot,label=below:] at (2.4,0.5) {};
\node[dot,label=above:] at (1.2,1.7) {};
\node[dot,label=right:] at (3.2,1.3) {};

\node[dot,label=below:] at (0.2,-0.8) {};
\node[dot,label=below:] at (1.5,-1.2) {};

\node at (1.6,-1.8) {\small (a) Point set $P$};

% =========================================================
% (b) A CPNCS solution
% =========================================================
\begin{scope}[xshift=5.4cm]

% Solution segments: ab, cd, pq, and bs
\draw[segment,segpurple]   (0,0.5) -- (2.4,0.5);     % ab
\draw[segment,segblue] (0.2,-0.8) -- (1.5,-1.2); % cd
\draw[segment,seggreen]  (1.2,0.5) -- (1.2,1.7);   % pq
\draw[segment,segorange] (2.4,0.5) -- (3.2,1.3);   % bs

\node[dot,label=below:] at (0,0.5) {};
\node[dot,label=below:] at (1.2,0.5) {};
\node[dot,label=below:] at (2.4,0.5) {};
\node[dot,label=above:] at (1.2,1.7) {};
\node[dot,label=right:] at (3.2,1.3) {};

\node[dot,label=below:] at (0.2,-0.8) {};
\node[dot,label=below:] at (1.5,-1.2) {};

\node at (1.6,-1.8) {\small (b) A CPNCS solution};

\end{scope}
\end{tikzpicture}

\caption{An instance of the {\sc CPNCS}   with $k=4$ (a) and a solution (b).}
\label{fig:cpncs-example}
\end{figure}

Covering a finite point set by a small number of simple geometric objects is a fundamental theme in computational geometry.  A basic example is {\sc Point Line Cover}: given a finite point set $P\subseteq\mathbb{R}^2$ and a positive integer $k$, decide whether $P$ can be covered by at most $k$ lines \cite{DBLP:journals/orl/MegiddoT82}.  Replacing lines by bounded segments does not change the problem when intersections are unrestricted: every segment has a supporting line, and every selected line can be shortened to the span of the input points assigned to it.  The problem changes substantially, however, when the covering segments are required to be non-crossing; two line segments are non-crossing if their relative interiors are disjoint.  A segment can obstruct several other supporting lines, and the points on a {\em rich line} (a line containing at least $k+1$ input points) may have to be covered by several separated segments.  A \emph{non-crossing segment cover} of $P$ is a family of pairwise non-crossing segments whose union contains $P$ (see \Cref{fig:cpncs-example}).  We study the following problem, introduced by Grelier~\cite{DBLP:conf/compgeom/Grelier22,grelier2019hardness}.

\begin{tcolorbox}[
    enhanced,
    title={\color{black}\textsc{Covering Points with Non-Crossing Segments (CPNCS)}},
    colback=white,
    boxrule=0.4pt,
    attach boxed title to top center={
        xshift=-.5cm,yshift*=-2.5mm
    },
    boxed title style={
        size=small,
        frame hidden,
        colback=white
    }
]
\vspace{-1mm}
\textbf{Input:} A finite point set $P\subseteq\mathbb{R}^{2}$, and  $k \in \mathbb{N}$.

\textbf{Question:} Can $P$ be covered by   $ \leq k$ pairwise non-crossing segments?
\end{tcolorbox}

% \begin{tcolorbox}[masterbox,title={{\sc Covering Points with Non-Crossing Segments (CPNCS)}}]
% \begin{tabular}{@{}lp{9.5cm}@{}}
% \textit{Input:}
% & A finite point set $P\subseteq\mathbb{R}^{2}$, and  $k \in \mathbb{N}$.\\[3pt]

% \textit{Question:}
% & Can $P$ be covered by   $ \leq k$ pairwise non-crossing segments?
% \end{tabular}
% \end{tcolorbox}

As shown in \Cref{sec:prelim}, requiring the endpoints to belong to $P$ does not change the optimum: after assigning multiply covered input points to suitable segments, every segment can be shortened to the span of its assigned points, and segments with no points assigned can be deleted. We write $\sco(P)$ for the minimum size of a non-crossing segment cover of $P$. Observe that $\lc(P)\leq\sco(P)$, where $\lc(P)$ is the minimum number of lines covering $P$, because the supporting lines of any segment cover form a line cover.

We also consider a supported version of the problem.  Here the input includes a family $\mathcal L=\{\ell_1,\ldots,\ell_t\}$ of prescribed lines, and every selected segment must be contained in a line of $\mathcal L$.  This variant isolates the effect of the line arrangement and provides a bridge between {\sc CPNCS} and {\sc Point Line Cover}.

\begin{tcolorbox}[
    enhanced,
    title={\color{black}\textsc{Supported CPNCS}},
    colback=white,
    boxrule=0.4pt,
    attach boxed title to top center={
        xshift=-3.5cm,yshift*=-2.5mm
    },
    boxed title style={
        size=small,
        frame hidden,
        colback=white
    }
]
\vspace{-1mm}
\textbf{Input:} A finite point set $P\subseteq\mathbb{R}^{2}$, a family $\mathcal L$ of
  $t$ lines, and  $k \in \mathbb{N}$.

\textbf{Question:} Can $P$ be covered by at most $k$ pairwise non-crossing line segments, each
  contained in a line of $\mathcal L$?
\end{tcolorbox}

% \begin{tcolorbox}[variantbox,title={{\sc Supported CPNCS}}]
% \begin{tabular}{@{}lp{9cm}@{}}
% \textit{Input:}
% & A finite point set $P\subseteq\mathbb{R}^{2}$, a family $\mathcal L$ of
%   $t$ lines, and  $k \in \mathbb{N}$.\\[3pt]

% \textit{Question:}
% & Can $P$ be covered by at most $k$ pairwise non-crossing line segments, each
%   contained in a line of $\mathcal L$?
% \end{tabular}
% \end{tcolorbox}

\paragraph{Related work.}
{\sc Point Line Cover} is {\sf NP}-hard \cite{DBLP:journals/orl/MegiddoT82}, but is fixed-parameter tractable when parameterized by the solution size.  Langerman and Morin \cite{DBLP:journals/dcg/LangermanM05} established this in a general framework that also covers hyperplanes in fixed-dimensional spaces.  The standard rich-line rule yields a kernel with at most $k^2$ points: any line containing more than $k$ uncovered points is forced.  This bound is essentially optimal, as an $\bigO(k^{2-\varepsilon})$-point kernel would imply $\mathsf{coNP}\subseteq\mathsf{NP/poly}$ \cite{DBLP:journals/talg/KratschPR16}.  When only horizontal and vertical lines are allowed, the problem reduces to {\sc Bipartite Vertex Cover} and is polynomial-time solvable.

Further work has addressed approximation and faster parameterized algorithms. Dumitrescu and Jiang studied the approximability of {\sc Point Line Cover} \cite{DBLP:journals/comgeo/DumitrescuJ15}; Afshani et al.\ used incidence bounds to obtain improved parameterized algorithms for covering points by lines, curves, and hyperplanes \cite{DBLP:conf/compgeom/AfshaniBDN16}; and Chen et al.\ developed faster rich-line detection and kernelization algorithms \cite{DBLP:conf/stacs/ChenHKX22}. Kowalska and Pilipczuk \cite{DBLP:conf/stacs/KowalskaP24} studied a closely related problem: covering points by segments from a given set.

{\sc CPNCS} arose from the minimum convex partition problem, for which Knauer and Spillner studied approximation algorithms \cite{DBLP:conf/swat/KnauerS06}, Magdalene Grantson and Christos Levcopoulos gave an {\sf FPT} algorithm parameterized by the hole vertices and/or the reflex vertices \cite{DBLP:conf/jcdcg/GrantsonL04}, and Spillner gave an {\sf FPT} algorithm parameterized by the number of points interior to the convex hull \cite{DBLP:journals/jda/Spillner08}, .  Grelier introduced {\sc CPNCS} and proved it {\sf NP}-hard, obtained an $\bigO(\log\mathrm{OPT})$-approximation, and gave an exact algorithm running in $\mathcal{O}(2^{k^2}k^{7k}+n^4\log n)$ \cite{DBLP:conf/compgeom/Grelier22,grelier2019hardness}.  The related {\sc Segment Cover} problem requires the selected segments to be pairwise non-intersecting, thereby forbidding even common endpoints. {\sf NP}-hardness of {\sc Segment Cover} follows from Grelier’s result and \Cref{lem:model-equivalence}; Garijo, M\'arquez, and Silveira \cite{DBLP:journals/corr/abs-2602-17294} gave an independent reduction that extends to guillotine cuts.  We show below that our results for {\sc CPNCS} also extend to {\sc Segment Cover}.

\paragraph{{\sc CPNCS} is equivalent to {\sc Segment Cover}.} Although {\sc CPNCS} allows two segments to touch at a point, its optimum is the same as that of {\sc Segment Cover}, where segments must be pairwise non-intersecting (\Cref{lem:model-equivalence}). Hence, {\sc CPNCS} and {\sc Segment Cover} have the same optimum, and results for one problem apply to the other.

\paragraph{Our results and methods.} For ease of exposition, the technical sections consider {\sc Segment Cover} and {\sc Supported Segment Cover}, whose solution segments are pairwise non-intersecting.  By \Cref{lem:model-equivalence}, every non-crossing solution can be transformed, without increasing its size or changing its supporting lines, into a non-intersecting one.  Thus all our results also hold for {\sc CPNCS} and {\sc Supported CPNCS}.

\begin{itemize}

\item \textbf{Parameterization by $k$ (\Cref{sec:seg}).} We improve Grelier's $\bigO(2^{k^2}k^{7k}+n^4\log n)$-time algorithm \cite{DBLP:conf/compgeom/Grelier22,grelier2019hardness} to $2^{\bigO(k\log k)}n^{\bigO(1)}$.  Every line containing at least $k+1$ points must support a solution segment; hence a \yes-instance has at most $k$ such rich lines and at most $k^2$ points outside them.  We construct $\bigO(k^3)$ candidate segments and enumerate adapted families of at most $k$ candidates within the claimed running time.

\item \textbf{Supported Segment Cover (\Cref{sec:restricted}).} For $t$ prescribed supporting lines, we obtain a kernel with $\bigO(t^2)$ points (\Cref{sec:kernel}) by retaining input intersection points and the extreme input points of every nonempty block.  We also give a $2^{\bigO(t)}(n+t)^{\bigO(1)}$-time sweep-line algorithm (\Cref{sec:fpt1}); its states record the subset of lines whose selected segments are currently active.  Parameterized by the solution size $k$, the supported problem is solvable in $2^{\bigO(k\log k)}(n+t)^{\bigO(1)}$ time (\Cref{sec:fpt2}).

\item \textbf{Parameterization by line cover number $\lc(P)$ (\Cref{sec:linecover}).} Writing $r=\lc(P)$ for the minimum number of lines covering $P$, we prove $r\leq \sco(P)\leq \frac{r(r+1)}{2}$. Together with our main algorithm, this gives an $r^{\bigO(r^2)}n^{\bigO(1)}$-time algorithm parameterized by $r$.

\end{itemize}

% Due to space constraints, proofs of statements marked with
% $\spadesuit$ appear in the appendix.

\section{Preliminaries}\label{sec:prelim}
\paragraph{Parameterized Complexity.} A \emph{parameterized problem} is a language $\Pi \subseteq \Sigma^{*} \times \mathbb{N}$, where $\Sigma$ is a finite fixed alphabet. For an instance $(I,k)\in \Sigma^{*} \times \mathbb{N}$, $k$ is called the \emph{parameter}.  A parameterized problem $\Pi$ is \emph{fixed-parameter tractable} (for short {\sf FPT}) if there exists an algorithm that decides whether $(I,k)\in \Pi$ in time $f(k)\cdot |I|^{\mathcal{O}(1)}$, where $f$ is a computable function depending only on $k$. Such an algorithm is called an \emph{{\sf FPT} algorithm}, and the class of all fixed-parameter tractable problems is denoted by {\sf FPT}. For a parameterized problem $\Pi$, a kernelization algorithm maps an instance $(I,k)$ in polynomial time to an equivalent instance $(I',k')$ whose encoding length $|I'|$ and parameter $k'$ are bounded by a computable function $g(k)$. The output $(I',k')$ is called a kernel. If $g(k)\in k^{\mathcal{O}(1)}$, then we say $\Pi$ admits a polynomial kernel. We refer to the monographs \cite{cygan2015parameterized,DBLP:series/mcs/DowneyF99,DBLP:books/ox/Niedermeier06} for a detailed study of parameterized complexity.

\smallskip

Let $P\subseteq\mathbb{R}^2$ be a finite set of $n$ points.  A \emph{$P$-segment} is the convex hull of one or two points of $P$; a one-point segment is \emph{degenerate}.  We set the relative interior of a one-point segment $\{p\}$ as $\relint(\{p\})=\{p\}$. Two segments are \emph{intersecting} if they have a nonempty intersection. Two segments \emph{cross} if their relative interiors intersect. Thus, for two non-crossing segments, endpoint contact is allowed, but overlapping collinear segments are crossing.

\begin{lemma}[Equivalence of Non-crossing and Non-intersecting] \label{lem:model-equivalence}
For every finite point set $P$, the minimum number of pairwise non-crossing segments covering $P$ equals the minimum number of pairwise non-intersecting segments covering $P$.  The same statement holds when all segments are required to be supported by a prescribed family of lines.
\end{lemma}
\begin{proof}
Every pairwise non-intersecting family is pairwise non-crossing, so one inequality is immediate.  For the converse, let $\mathcal S$ be a pairwise non-crossing segment cover of $P$.

Assign every point $p\in P$ to one segment of $\mathcal S$ containing it.  If $p$ lies in the relative interior of such a segment, assign $p$ to that segment; this segment is unique because the relative interiors of the members of $\mathcal S$ are pairwise non-intersecting.  If $p$ lies only at segment endpoints, assign it to an arbitrary segment containing it.

For every $s\in\mathcal S$, replace $s$ by the segment spanning the points assigned to it, and delete $s$ if no point is assigned to it.  The resulting segments still cover $P$, and their number does not increase.

We claim that they are pairwise non-intersecting.  Consider an intersection point $x$ of two original segments.  Since the segments are non-crossing, $x$ is an endpoint of at least one of them.  If $x\notin P$, every segment having $x$ as an endpoint retracts from $x$ during trimming.  If $x\in P$ and lies in the relative interior of one segment, then $x$ is assigned to that segment, and every segment having $x$ as an endpoint retracts from it.  Finally, if $x$ is an endpoint of every segment containing it, then it is assigned to only one of them, and all the others retract from $x$.  Hence, no intersection remains.

Each replacement segment is contained in its original segment and therefore has the same supporting line.  Thus, the transformation also preserves any prescribed family of supporting lines.
\lineqed 
\end{proof}

\begin{tcolorbox}[
    enhanced,
    title={\color{black}\textsc{\SC{}}},
    colback=white,
    boxrule=0.4pt,
    attach boxed title to top center={
        xshift=-3.5cm,yshift*=-2.5mm
    },
    boxed title style={
        size=small,
        frame hidden,
        colback=white
    }
]
\vspace{-1mm}
\textbf{Input:} A finite point set $P\subseteq\mathbb{R}^{2}$, and an integer $k$.

\textbf{Question:} Can $P$ be covered  by $ \leq k$ pairwise non-intersecting line segments?
\end{tcolorbox}

% \begin{tcolorbox}[masterbox,title={\sc \SC{}}]
% \begin{tabular}{@{}lp{10cm}@{}}
% \textit{Input:}
% & A finite point set $P\subseteq\mathbb{R}^{2}$, and an integer $k$.\\[3pt]

% \textit{Question:}
% & Can $P$ be covered  by $ \leq k$ pairwise non-intersecting line segments?
% \end{tabular}
% \end{tcolorbox}

\begin{figure}[t!]
\centering

\begin{tikzpicture}[
    dot/.style={circle,fill=black,inner sep=1.4pt},
    seg/.style={line width=0.8pt}
]

% =========================================================
% (a) Input point set
% =========================================================

% Group 1
\node[dot] at (0,1.4) {};
\node[dot] at (0.8,1.1) {};
\node[dot] at (1.6,0.8) {};

% Group 2
\node[dot] at (0.3,-0.3) {};
\node[dot] at (1.2,-0.3) {};
\node[dot] at (2.1,-0.3) {};

% Group 3
\node[dot] at (2.5,1.5) {};
\node[dot] at (3.0,0.8) {};
\node[dot] at (3.5,0.1) {};

\node at (1.75,-1.0) {\small (a) Point set $P$};

% =========================================================
% (b) Segment cover
% =========================================================
\begin{scope}[xshift=6.0cm]

% Covering segments
\draw[seg] (0,1.4) -- (1.6,0.8);
\draw[seg] (0.3,-0.3) -- (2.1,-0.3);
\draw[seg] (2.5,1.5) -- (3.5,0.1);

% Group 1
\node[dot] at (0,1.4) {};
\node[dot] at (0.8,1.1) {};
\node[dot] at (1.6,0.8) {};

% Group 2
\node[dot] at (0.3,-0.3) {};
\node[dot] at (1.2,-0.3) {};
\node[dot] at (2.1,-0.3) {};

% Group 3
\node[dot] at (2.5,1.5) {};
\node[dot] at (3.0,0.8) {};
\node[dot] at (3.5,0.1) {};

\node at (1.75,-1.0) {\small (b) A solution};

\end{scope}

\end{tikzpicture}

\caption{%
An instance of the {\sc Segment Cover}   with $k=3$ (a) and a solution (b).
}
\label{fig:segment-cover-example}
\end{figure}
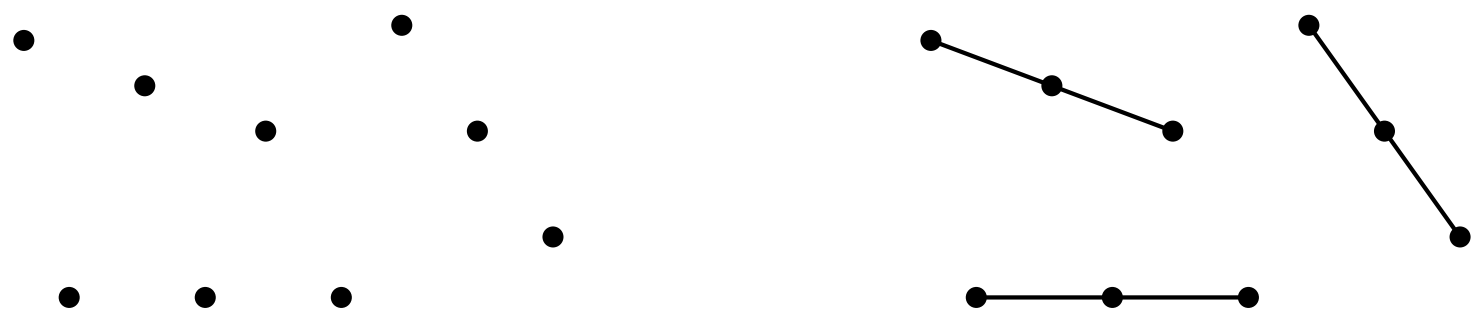

\begin{tcolorbox}[
    enhanced,
    title={\color{black}\textsc{Supported Segment Cover}},
    colback=white,
    boxrule=0.4pt,
    attach boxed title to top center={
        xshift=-2.5cm,yshift*=-2.5mm
    },
    boxed title style={
        size=small,
        frame hidden,
        colback=white
    }
]
\vspace{-1mm}
\textbf{Input:} A finite point set $P\subseteq\mathbb{R}^{2}$, a family $\lee$ of $t$ lines, and an integer $k$.

\textbf{Question:} Can $P$ be covered  by $ \leq k$ pairwise non-intersecting line segments each supported by a line of $\lee$?
\end{tcolorbox}

% \begin{tcolorbox}[variantbox,title={\sc Supported Segment Cover}]
% \begin{tabular}{@{}lp{10cm}@{}}
% \textit{Input:}
% & A finite point set $P\subseteq\mathbb{R}^{2}$, a family $\lee$ of $t$ lines, and an integer $k$.\\[3pt]

% \textit{Question:}
% & Can $P$ be covered  by $ \leq k$ pairwise non-intersecting line segments each supported by a line of $\lee$?
% \end{tabular}
% \end{tcolorbox}
By Lemma~\ref{lem:model-equivalence}, the two formulations are equivalent for our purposes. {\textbf{Therefore, for simplicity of exposition, throughout the remainder of the paper we present all algorithms in terms of {\sc Segment Cover} and {\sc Supported Segment Cover}, rather than {\sc CPNCS} and {\sc Supported CPNCS}}. The same algorithms and bounds apply to the corresponding {\sc CPNCS} variants, since every non-crossing solution can be converted in polynomial time into a pairwise non-intersecting solution of no greater size.

\section{Segment Cover}\label{sec:seg}

Requiring endpoints in $P$ does not change the optimum: any segment can be shortened to the convex hull of the points of $P$ that it covers, and shortening creates no new intersection. Segments covering no input point can be discarded.

For a line $\ell$, write $P_\ell=P\cap\ell$, ordered along $\ell$.  If $P_\ell\ne\emptyset$, let $e_\ell=\operatorname{conv}(P_\ell)$.  A line is \emph{rich} (with respect to $k$) if $|P_\ell|\ge k+1$.  Let $\mathcal R$ be the family of rich lines, let $h=|\mathcal R|$, and call $Q=P\setminus\bigcup_{\ell\in\mathcal R}\ell$ the set of \emph{poor points}.

\begin{observation}\label{obs:two}
If a $P$-segment contains two points of $P_\ell$, then it is supported by $\ell$ and is contained in $e_\ell$.
\end{observation}

\begin{lemma}\label{lem:rich}
If $P$ has a segment cover of size at most $k$, then every rich line supports a member of the cover, $h\le k$, and $|Q|\le k^2$.  The rich lines and the poor points can be computed in $n^{\bigO(1)}$ time.
\end{lemma}
\begin{proof}
If no chosen segment is supported by a rich line $\ell$, each of the at most $k$ segments of the cover meets $\ell$ in at most one point, and therefore the cover contains at most $k<|P_\ell|$ points of $\ell$.  Hence, every rich line needs a distinct supporting segment and $h\le k$.

No line contains more than $k$ poor points: a line containing $k+1$ of them would itself be rich, contrary to the definition of $Q$.  Each of the at most $k$ covering segments therefore covers at most $k$ poor points, so $|Q|\le k^2$.

By exhaustive search, we can find rich lines and poor points in $n^{\bigO(1)}$ time.
\lineqed
\end{proof}
We henceforth reject when $h>k$ or $|Q|>k^2$.  Notice also that if $h=0$, a \yes-instance has at most $k^2$ points.

Fix a segment cover $\see$ of size at most $k$.  The following notion is adapted from Grelier's rich-line branching~\cite{DBLP:conf/compgeom/Grelier22,grelier2019hardness}.

\begin{definition}[Adapted family]\label{adapted}
{\em A family $L$ of $P$-segments is \emph{adapted to $\see$} if
\begin{itemize}
  \item[(a)] every $s\in L$ contains at least $k+1$ input points;
  \item[(b)] if $s\in L$ and $\sigma\in \see$ share at least two input points, then $\sigma\subseteq s$;
  \item[(c)] every $\sigma\in \see$ containing at least $k+1$ input points is contained in some member of $L$.
\end{itemize}}
\end{definition}
\begin{lemma}\label{reverse}
    Suppose a family $L$ is adapted to a cover $\see$. Then for each $s\in L$, there exists nondegenerate $\sigma_s\in \see$ such that $\sigma_s \subseteq s$.
\end{lemma}
\begin{proof}
    Let $s\in L$. Using condition (a), $s$ contains at least $k+1$ input points. Now, if each $\sigma \in \see$ shares at most one input point with $s$, then $\see$ cannot cover all the input points of $s$, which is a contradiction. Thus there is a $\sigma_s \in \see$ such that $|\sigma\cap s\cap  P|\ge 2$. Using condition (b), $\sigma_s \subseteq s$.
     \lineqed 
\end{proof}

\begin{figure}[ht!]
\centering

\begin{tikzpicture}[
    line cap=round,
    line join=round,
    mainline/.style={draw=black, line width=0.8pt},
    point/.style={circle, fill=black, inner sep=1.5pt},
    plabel/.style={font=\small},
    linelabel/.style={font=\small\itshape}
]

% ============================================================
% Case (i)
% ============================================================
\begin{scope}[xshift=0cm]

    \coordinate (x1) at (0,0);
    \coordinate (u1) at (1.65,0.62);
    \coordinate (p1) at (2.10,0.79);
    \coordinate (v1) at (2.65,1.00);
    \coordinate (y1) at (4.85,1.83);

    \coordinate (a1) at (0,1.83);
    \coordinate (b1) at (3.65,0);

    % Lines
    \draw[mainline] (x1) -- (y1);
    \draw[mainline] (a1) -- (b1);

    % Points
    \node[point] at (x1) {};
    \node[point] at (u1) {};
    \node[point] at (p1) {};
    \node[point] at (v1) {};
    \node[point] at (y1) {};

    % Point labels
    \node[plabel, below left=1pt and 1pt of x1] {$x$};
    \node[plabel, below=2pt of u1] {$u$};
    \node[plabel, above=2pt of p1] {$p$};
    \node[plabel, below=2pt of v1] {$v$};
    \node[plabel, below=2pt of y1] {$y$};

    % Line labels
    \node[linelabel, above left=1pt]
        at ($(x1)!0.22!(u1)$) {$s$};

    \node[linelabel, above right=1pt]
        at ($(a1)!0.10!(p1)$) {$s'$};

    % Panel label
    \node at (2.35,-0.65) {(i)};

\end{scope}

% ============================================================
% Case (ii)
% ============================================================
\begin{scope}[xshift=7.1cm]

    \coordinate (x2) at (0,0);
    \coordinate (u2) at (1.65,0.62);
    \coordinate (p2) at (2.10,0.79);

    \coordinate (a2) at (0,1.83);
    \coordinate (b2) at (3.65,0);

    % Lines
    \draw[mainline] (x2) -- (p2);
    \draw[mainline] (a2) -- (b2);

    % Points
    \node[point] at (x2) {};
    \node[point] at (u2) {};
    \node[point] at (p2) {};

    % Point labels
    \node[plabel, below left=1pt and 1pt of x2] {$x$};
    \node[plabel, below=2pt of u2] {$u$};
    \node[plabel, above=2pt of p2] {$p$};

    % Line labels
    \node[linelabel, above left=1pt]
        at ($(x2)!0.22!(u2)$) {$s$};

    \node[linelabel, above right=1pt]
        at ($(a2)!0.10!(p2)$) {$s'$};

    % Panel label
    \node at (1.95,-0.65) {(ii)};

\end{scope}

\end{tikzpicture}

\caption{
Case (i):
$g_s(s')=\overline{uv}$,
$E_s(s')=\{u,v\}$, and
$F_s(s')=\{\overline{xu},\overline{vy}\}$.
Case (ii):
$g_s(s')=\overline{up}$,
$E_s(s')=\{u,p\}$, and
$F_s(s')=\{\overline{xu}\}$.
}
\label{enclosing}

\end{figure}
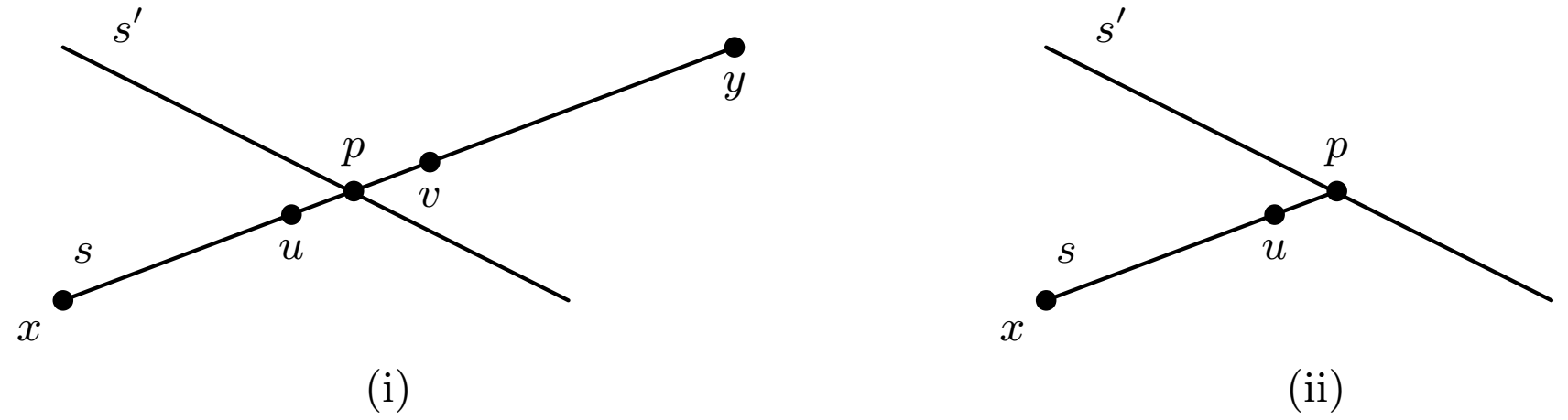

For two intersecting $P$-segments (non-parallel) $s,s'$ with intersection $p$, we define $g_s(s')$ to be the shortest $P$-segment contained in $s$ such that $p\in g_s(s')$ and $ g_s(s')\neq \{p\}$. Let $E_s(s')$ be the set of the two endpoints of $g_s(s')$. Let $F_s(s')$ be the set of all connected components of the closure of $s\setminus g_s(s')$. Clearly, $F_s(s')$ consists of at most two closed segments. See \Cref{enclosing} for an illustration. Let $T_s(s')$ denote the subset of $F_s(s')$ consisting of segments that contain at least $k+1$ input points.

For $\ell\in \ree$, define $B_\ell$ to consist of the two endpoints of $e_\ell$, and the points in $E_{e_\ell}(e_{\ell'})$ for each $\ell'\in \ree$ with $e_{\ell}\cap e_{\ell'}\neq \emptyset, ~\ell'\neq \ell$. Then $|B_\ell|\le2+2(h-1)\le2k$. Let $\iee_\ell$ be the $P$-segments on $\ell$ whose endpoints belong to $B_\ell$ and that contain at least $k+1$ points of $P$, and put $\iee=\bigcup_{\ell\in\mathcal R}\iee_\ell$.  We have
$|\iee|\le h\binom{2h}{2}=\bigO(k^3).$

\begin{lemma}\label{lem:adapted-exists}
If $\see$ is a segment cover of size at most $k$, then there is a pairwise non-intersecting family $L\subseteq \iee$ adapted to $\see$, with $|L|\le k$.

\end{lemma}
\begin{proof}
Start with $L=\{e_\ell:\ell\in\mathcal R\}$ which is adapted to $\see$. Indeed, conditions (a)--(c) hold: (a) is the definition of richness, while (b) and (c) follow from \Cref{obs:two}. For each $s,s'\in L$ such that $s$ and $s'$ are intersecting, we do the following:  Let the intersection point of $s$ and $s'$ be $p$.
\begin{description}
    \item[Case I ($p\in P$).]  Let $e\in \see$ cover $p$. Now, $e$ cannot be collinear with both $s$ and $s'$. If $e$ is not collinear with $s$ then we remove $s$ from $L$ and add each $l\in T_s(s')$ to $L$. Next, we do a similar process if $e$ is not collinear with $s'$ as we did for $s$.

\smallskip 
    \item[Case II ($p\notin P$).] Observe that either the interior of $g_s(s')$ is contained in some segment of $\see$, or all segments of $\see$ are disjoint from it. Now $g_s(s')$ and $g_{s'}(s)$ cannot both be contained in members of $\see$, since $g_s(s')$ and $g_{s'}(s)$ have non-empty intersection. If $g_s(s')$ is not contained in any segment of $\see$, remove $s$ from $L$ and add each $l\in T_s(s')$. We do a similar process for $s'$ if $g_{s'}(s)$ is not contained in any segment of the covering.
\end{description}

\begin{clm} \label{claim:each}
    After each step of updating $L$ (i.e., removing $s$ and inserting segments of $T_s(s')$ according to the rules described above), the updated $L$ is adapted to $\see$.
\end{clm}

\begin{claimproof}
    Here, our assumption is $s,s'$ (two elements of $L$ before this step begins) intersect at $p$ and the previous $L$ is adapted to $\see$. After this step, $s$ has been removed, and $l\in T_s(s')$ has been inserted into $L$ according to the rules described above. For the new $L$, condition (a) is preserved by construction. For (b), suppose a solution segment $\sigma$ shares two input points with $l\in T_s (s')$. Thus $|s\cap \sigma|\geq 2$ because $l\subseteq s$. Since the previous $L$ is adapted to $\see$, we have $\sigma\subseteq s$. Now, if $\sigma$ is not contained in $l$ then either $\sigma$ covers $p$ (for $p\in P$) or $g_s(s')\subseteq \sigma$ (for $p\notin P$). Both contradict the process of updating $P$. Thus $\sigma\subseteq l$, and condition (b) holds for the updated $L$. For (c), $\sigma~(\in \see)$ containing at least $k+1$ points and formerly contained in $s$ either cannot contain $p$ (for $p\in P$), or cannot contain $g_s(s')$ (for $p\notin P$) and is therefore contained in one of the two segments of $F_s(s')$; this segment is inserted in $L$ because it already contains at least $k+1$ points.
\hfill$\diamond$
\end{claimproof}

The operation resolves the unique intersection of the two supporting rich lines and never reintroduces it.  Hence, it terminates after resolving all the pairwise intersection points, and the final family is pairwise non-intersecting.

Using \Cref{reverse}, for each $s\in L$, there is a nondegenerate $\sigma_s\in \see$ with $\sigma_s\subseteq s$. Distinct members of the pairwise non-intersecting family $L$ contain distinct nondegenerate members of $\see$.  Therefore $|L|\le|\see|\le k$

Let $s\subset \ell~(\ell\in \ree)$, then the endpoints of each segments of $F_s(s')$ are in $B_{\ell}$. Thus $l$ has been inserted to $L$ implies that $l\in \iee_{\ell}$. Thus $L\subset \iee$.
\lineqed
\end{proof}

Any subset of $\iee$ of size at most $k$ is called a \emph{candidate family}.
\begin{corollary}\label{enumeration}
There are $2^{\bigO(k\log k)}$ candidate families, and they can be enumerated and tested for pairwise intersection within $2^{\bigO(k\log k)}n^{\bigO(1)}$ time.
\end{corollary}
\begin{proof}
Writing $N=|\iee|=\bigO(k^3)$,
$\sum_{i=0}^k\binom Ni\le(k+1)\left(\frac{eN}{k}\right)^k =2^{\bigO(k\log k)}.$
Each segment has two input endpoints, so both coverage and intersection tests require a polynomial number of orientation tests.
\lineqed
\end{proof}

For a current candidate family $L$ and a partial solution $D$, define $F_{_L}=P\setminus\bigcup_{s\in L} s$ and $R_{_{L,D}}=P\setminus\left(\bigcup _{s\in L} s\cup\bigcup_{\sigma\in D} \sigma\right).$
Thus $F_{_{L}}$ consists of points free of the candidate segments, whereas $R_{_{L,D}}$ consists of points not yet covered at all, and $R_{_{L,D}}\subseteq F_{_{L}}$.

\begin{lemma}\label{progress} Suppose $L$ is adapted to a solution $\see$ of size at most $k$, $D\subseteq \see$, and $D\cup L$ is pairwise non-intersecting. Then $|F_{_{L}}|\le k^2$.  Moreover, if $|D|+|L|+|R_{_{L,D}}|>k$, some segment of $\see\setminus D$ contains at least two points of $R_{_{L,D}}$.
\end{lemma}
\begin{proof}
If a segment of $\see$ contains a point of $F_{_{L}}$, it contains at most $k$ input points; otherwise, condition (c) would place it inside a member of $L$.  The $k$ solution segments therefore cover at most $k^2$ points of $F_{_{L}}$.

For every $s\in L$, \Cref{reverse} supplies a distinct segment $\sigma_s\in \see$ contained in $s$. Let $\Omega=\{\sigma_s|s\in L\}$, clearly $|\Omega|=|L|$. Now $\Omega\cap D=\emptyset$, because $D$ is non-intersecting with $L$. The segments of $\see\setminus (D\cup \Omega)$ cover $R_{_{L,D}}$.  If each segment of $\see\setminus (D\cup \Omega)$ covered at most one point of $R_{_{L,D}}$, then $|\see\setminus (D\cup \Omega)|\geq |R_{_{L,D}}|$. Thus, $|\see|=|D|+|\Omega|+|\see\setminus (D\cup \Omega)|\ge|D|+|L|+|R_{_{L,D}}|>k$, a contradiction.
\lineqed
\end{proof}

\begin{lemma}\label{endpoint}
Under the assumptions of \Cref{progress}, let $\sigma\in \see\setminus D$ contain at least two points of $R_{_{L,D}}$, and let $\tau$ be the shortest subsegment of $\sigma$ containing $F_{_{L}}\cap\sigma$.  Each endpoint of $\sigma$ is either an endpoint of $\tau$ or an input point of $\sigma$ lying in some member of $L$.  Once $\tau$ is fixed, $\sigma$ has $k^{\bigO(1)}$ possible endpoint pairs that can be generated from $L$.
\end{lemma}
\begin{proof}
The set $F_{_{L}}\cap\sigma$ contains the two residual points, so $\tau$ is a nondegenerate $P$-segment.  An endpoint $q$ of $\sigma$ lying outside $\tau$ is not in $F_{_{L}}$, by the definition of $\tau$, and hence belongs to some $s\in L$.  Since $\sigma$ contains a point of $F_{_{L}}$, it is not contained in $s$; condition (b) therefore gives $|P\cap\sigma\cap s|\le1$. 

Now, suppose we have fixed a $\tau$ for $\sigma$. For $s\in L$ not collinear with $\sigma$, the only candidate for an endpoint of $\sigma$ is the intersection of their supporting lines, provided it belongs to $P\cap s$. Now, if $s$ is collinear with $\sigma$ and $\sigma$ has an endpoint in $s$, then this endpoint must be the endpoint of $s$, which is nearest to $\sigma$. Otherwise, entering further into $s$ would place at least two input points of $s$ on $\sigma$ and since $L$ is adapted to $\see$, condition (b) gives $\sigma\subseteq s$ which is a contradiction since $\tau$ is the smallest subsegment of $\sigma$ containing $F_{_L}\cap\sigma$ and $\tau$ is nonempty. So, for a collinear member, the only possible candidates are its two extreme input points. Together with the endpoints of $\tau$, this gives at most $2|L|+2\le2k+2$ candidates for each endpoint.
\lineqed
\end{proof}
\begin{lemma}\label{trim}
In the setting of \Cref{progress}, let $\sigma\in \see\setminus D$ contain a point of $F_{_L}$.  Replace every member $s$ of $L$ that intersects $\sigma$ by $T_s(\sigma)$.  The new family is adapted to $\see$, remains pairwise non-intersecting, and is non-intersecting with $D\cup\{\sigma\}$.
\end{lemma}
\begin{proof}
Updating $L$ only shortens candidate segments, so it creates no intersection among them or with $D$. The only thing to prove is that the new family $L'$ is adapted to $\see$. Condition (a) holds by the construction. For (b), let $\sigma'\in \see$ shares at least two input points with $s'\in L'$. Let $s'\in T_s(\sigma)$ for some $s\in L$. Since $L$ is adapted to $\see$ and $s'\subseteq s$, we have $\sigma'\subseteq s$. Now if $\sigma'$ is not contained in $s'$ then by construction $\sigma'$ must intersect $\sigma$ which is a contradiction since $\sigma, \sigma'\in \see$ and $\sigma\neq \sigma'$. For (c), let $\rho\in \see$ contains at least $k+1$ input points. Since $L$ is adapted to $\see$, $\rho\subseteq s$ for some $s\in L$. Then $\rho$ must be contained in some $s'\in T_s(\sigma)$, because otherwise $\rho$ intersects $\sigma$ which is a contradiction ($\rho\ne\sigma$ because $\sigma$ contains a point outside $L$).
\lineqed
\end{proof}
\begin{lemma}\label{completion}
Given a pairwise non-intersecting family $L$ of at most $k$ $P$-segments, one can in $k^{\bigO(k)}n^{\bigO(1)}$ time either output a segment cover of size at most $k$ or correctly report no such cover exists. If $L$ is adapted to some segment cover $\see$ of size at most $k$, the procedure outputs $\see$.
\end{lemma}
\begin{proof}
Maintain a partial family $D$, initially empty.  At every search node:
\begin{enumerate}
  \item Reject the node if $|L|>k$, $|D|>k$, or $|F_{_L}|>k^2$.
  \item If $|D|+|L|+|R_{_{L,D}}|\le k$, output $D\cup L\cup\{\{p\}:p\in R_{_{L,D}}\}$.  This family is pairwise non-intersecting: points of $R_{_{L,D}}$ lie in no member of $D\cup L$.
  \item Otherwise enumerate a pair of points of $F_{_{L}}$ as the endpoints of $\tau$.  Require that $\tau$ contain at least two points of $R_{_{L,D}}$.  Generate endpoint candidates for $\sigma$ as in \Cref{endpoint}, retain only $P$-segments containing $\tau$ and non-intersecting with $D$, and branch on all such choices.
  \item For each possible $\sigma$ enumerated in step 3, we open a new search node and update $ D$ and $ L$ as follows: add the selected $\sigma$ to $D$, update $L$ by removing $s$ and adding each $l\in T_s(\sigma)$ to $L$ if $\sigma$ intersects $s$.
\end{enumerate}

There are at most $\binom{k^2}{2}=\bigO(k^4)$ choices for $\tau$ and $\bigO(k^2)$ endpoint pairs for $\sigma$.  The depth is at most $k$, so the search has $k^{\bigO(k)}$ nodes; all geometric operations are polynomial in $n$.

Soundness follows from the explicit check and output in step (2).  For completeness, fix a family $L$ adapted to $\see$.  \Cref{progress} supplies a solution segment $\sigma$ whenever step (2) does not apply.  The correct $\tau$ and endpoints occur among the branches by \Cref{endpoint}, and \Cref{trim} preserves the invariant. Following these branches keeps $D\subseteq \see$, so the search eventually succeeds.
\lineqed
\end{proof}

\begin{theo}\label{thm:main}
\SC~can be solved in $k^{\bigO(k)}n^{\bigO(1)}$ time.  
\end{theo}
\begin{proof}
Apply \Cref{lem:rich}; reject if its necessary conditions fail.  Construct $\iee$, enumerate every pairwise non-intersecting subset of size at most $k$ as in \Cref{enumeration}, and run \Cref{completion} on each. If a solution of size at most $k$ exists, \Cref{lem:adapted-exists} ensures that one enumerated family is adapted to it, and completion succeeds. Every accepted branch explicitly returns a valid cover. Finally, using \Cref{enumeration} and \Cref{completion}, the runtime of this algorithm is $k^{\bigO(k)}n^{\bigO(1)}$.
\lineqed
\end{proof}

\section{Supported Segment Cover}\label{sec:restricted}

In \LRSC, the input additionally contains a family $\lee=\{\ell_1,\ldots,\ell_t\}$ of distinct lines, and every chosen segment must be supported by a member of $\lee$; a segment is \emph{supported by} a line $\ell$ if it is contained in $\ell$. Duplicate lines can be removed in polynomial time. It is easy to verify that \LRSC~is {\sf NP}-hard as it is a generalization of {\sc Segment Cover}. Hence we study this problem in parameterized realm with respect to several parameters such as $t, k$.

\subsection{Kernelization parameterized by  $t$}\label{sec:kernel}
We first apply the following immediate rule.

\begin{reductionrule}\label{rr1}
    If a point of $P$ lies on no line of $\mathcal L$, return a trivial {\tt No}-instance.
\end{reductionrule}

\noindent For a \yes-instance, we assume that $P\subseteq\bigcup\mathcal L$. For each $i$, let
\[
  X_i=\{\ell_i\cap\ell_j:j\ne i\text{ and }\ell_i
       \text{ is not parallel to }\ell_j\}.
\]
The connected components of $\ell_i\setminus X_i$ are called the \emph{atoms} of $\ell_i$.  Since $|X_i|\le t-1$, every line has at most $t$ atoms. For every atom containing input points, define its \emph{block} as the convex hull of those points.  

\begin{reductionrule}\label{rr2}
    Construct $P'\subseteq P$ as follows.  For every $i$,
\begin{enumerate}
  \item include every point of $P\cap X_i$; and
  \item in every nonempty block $B$ of $\ell_i$, include the endpoints of $B$.
\end{enumerate}
\end{reductionrule}

\noindent \Cref{rr1} retains all the intersection points, so there are at most $\binom{t}{2}$ points in $P'$ retained by applying \Cref{rr1} on all the lines. For each $i$, there are at most $t$ atoms on $\ell_i$, from which \Cref{rr2} retains at most $2t$ points. Thus $|P'|\le t(2t)+\binom{t}{2}=2t^2+\binom{t}{2}=\bigO(t^2).$ 

\begin{theo}\label{thm:lr-kernel}
\LRSC~admits a kernel  with $\mathcal{O}(t^2)$ points.
\end{theo}
\begin{proof}
    The forward direction is trivial. Let $\see$ be a solution of $(P,k,\lee)$. Then $\see$ is a solution for $(P',k,\lee)$ since $P'\subseteq P$.

    For the other direction, let $(P',k,\lee)$ is a \yes-instance. Let $\see'$ be a solution for $(P',k,\lee)$ with minimum cardinality. Assume that $p\in P$ has not been covered by any segment of $\see'$. Clearly $p\notin P'$. Let $p$ lie in the interior of the block $B$. $p$ cannot be an endpoint of $B$ since the endpoints of $B$ are in $P'$. Let $u$ and $v$ be two endpoints of $B$, and let $s_u, s_v$ be two segments of $\see'$ covering $u$ and $v$ respectively. If $s_u=s_v$, then $B\subseteq s_u$, and hence $p$ is covered, a contradiction. Thus, assume that $s_u\neq s_v$. $B$ is contained in exactly one line of $\lee$; denote this line be $\ell$. Since neither $u$ nor $v$ lies on any line of $\lee$ other than $\ell$, and the supporting lines of $s_u$ and $s_v$ are contained in $\lee$, $\ell$ is the supporting line of both $s_u$ and $s_v$. No other line of $\lee$ intersects $\ell$ at any point of $B$, so no segment of $\see'$ other than $s_u$ and $s_v$ intersects $B$. Define $s:= s_u\cup B\cup s_v$. Clearly $\see'':=(\see'\setminus \{s_u,s_v\})\cup\{s\}$ is a pairwise non-intersecting family of segments which covers $P'$ and supported by lines of $\lee$. Furthermore, $\bigl|\see''\bigr|<\bigl|\see'\bigr|$, contradicting the minimality of $\see'$. Thus $\see'$ must cover every point of $P$. Hence $(P,k,\lee)$ is a \yes-instance.\lineqed
\end{proof}

Next, we provide an \fpt-algorithm for \LRSC~when parameterized by $t$, later we obtain an \fpt-algorithm for \LRSC~when parameterized by $k$.

\subsection{FPT algorithms parameterized by $t$}\label{sec:fpt1}

We first give a simple $2^{\bigO(t^2)}(n+t)^{\bigO(1)}$-time algorithm for \LRSC{}.  For every pair of nonparallel lines in $\mathcal L$, we guess one of the two lines that is forbidden from containing their intersection.  There are at most $2^{\binom{t}{2}}$ such choices, which we call \emph{signatures}.  For a fixed signature, deleting the forbidden intersection points divides each line into at most $t$ open segments. Every nonempty open segment is represented by the canonical segment spanning all input points in that block.  These canonical segments are pairwise non-intersecting, and a solution compatible with the signature exists if and only if they cover $P$ and their number is at most $k$.  Thus each signature can be tested in polynomial time, giving a total running time of $ 2^{\binom{t}{2}}(n+t)^{\bigO(1)} =2^{\bigO(t^2)}(n+t)^{\bigO(1)}.$

\medskip 
Now we give an improved algorithm with running time $2^{\bigO(t)}n^{\bigO(1)}$ using a sweep-line dynamic programming. Let $\OPT(P,\lee)$ denote the minimum number of pairwise non-intersecting segments, each supported by a line $\ell\in\lee$, whose union covers all points of $P$.
 
Let $\Lambda$ be the set of pairs of indices ${i,j}$ such that the corresponding lines $l_i$ and $l_j$ are nonparallel, and for $\{i,j\}\in\Lambda$ write $x_{ij}=\ell_i\cap\ell_j$. Let $X:=P\cup\{x_{ij}\;|\;\{i,j\}\in\Lambda\}$ be the set of \emph{events}. Rotate the coordinate system so that no line of $\lee$ is vertical and distinct events have distinct $x$-coordinates. Only finitely many directions are excluded, one for each line of $\lee$ and one for each pair of events, so such a rotation exists. Rotation preserves the value of $\OPT(P,\lee)$. Write $x(q)$ for the $x$-coordinate of a point $q$.

Let $e_1,\dots,e_m$ be the events in increasing order of $x$-coordinate, so that $ m \;\le\; |P|+\binom{t}{2} \;=\; \bigO(n+t^2). $ Choose $\gamma_0<x(e_1)$, then $x(e_i)<\gamma_i<x(e_{i+1})$ for $1\le i\le m-1$, and finally $\gamma_m>x(e_m)$; let $\Gamma_i$ be the vertical line with $x$-coordinate $\gamma_i$. For $i\in[m]$ let $ R_i \;:=\; \{\,\ell\in\lee : e_i\in\ell\,\} $ be the set of input lines through $e_i$. Call $e_i$ an \emph{input-point event} when $e_i\in P$, and an \emph{ordinary event} otherwise. A segment $s$ with endpoints $u$ and $v$, where $x(u)\le x(v)$, is called \emph{open at $\Gamma_i$} when $x(u)<\gamma_i<x(v)$. A line $\ell\in\lee$ is \emph{open at $\Gamma_i$} in a family $\see$ when some member of $\see$ contained in $\ell$ is open at $\Gamma_i$.

A solution $\see$ of size $\OPT(P,\lee)$ is called \emph{canonical} if every non-degenerate member of $\see$ is a $P$-segment, and every degenerate member containing a point $p\in P$ lie inside a disc centered at $p$ that intersects neither any $\Gamma_i$ nor any other member of $\see$, and contains no point of $X\setminus{p}$. We may assume, without loss of generality, that every solution of size $\OPT(P,\lee)$ is canonical.

Now we describe the sweep by a combinatorial object, which makes the dynamic programming a plain shortest-path computation.

\begin{definition}[Trace]
\label{def:trace}
{\em A \emph{trace} is a pair $\tau=(S_0,\dots,S_m;\mathtt{ow})$, where each $S_i$ is a subset of $\lee$ and $\mathtt{ow}$ assigns to every index $i$ with $e_i\in P$ a line $\mathtt{ow}(i)\in R_i$, called the \emph{owner} of $e_i$, subject to three rules.
\begin{description}
  \item[\textnormal{(T0)}] $S_0=S_m=\emptyset$.
  \item[\textnormal{(T1)}] If $e_i\notin P$, then $S_i=S_{i-1}$ and
        $|S_{i-1}\cap R_i|\le 1$.
  \item[\textnormal{(T2)}] If $e_i\in P$ and $\ell=\mathtt{ow}(i)$, then
        $S_i\setminus R_i=S_{i-1}\setminus R_i$,
        $S_{i-1}\cap R_i\subseteq\{\ell\}$ and $S_i\cap R_i\subseteq\{\ell\}$.
\end{description}
The \emph{cost} of $\tau$ is $ c(\tau):=\bigl|\{i\,|\, e_i\in P \text{ and } \mathtt{ow}(i)\notin S_{i-1}\}\bigr|$.}
\end{definition}

\begin{lemma}
\label{thm:trace}
Traces exist, and $\min_\tau c(\tau)=\OPT(P,\lee)$ when $P\subset \cup_{\ell\in \lee}\ell$.
\end{lemma}
\begin{proof}
We start with showing that every canonical solution yields a trace of the same cost.

Let $\see$ be canonical of size $\OPT(P,\lee)$. For $e_i\in P$, let $\mathtt{ow}(i)$ be the supporting line of the member containing $e_i$. Clearly $\mathtt{ow}(i)\in R_i$. Put $S_i:=\{\ell\in\lee\,|\,\ell \text{ is open at } \Gamma_i \text{ in } \see\}$.

Rule (T0) holds because no member of $\see$ is open at $\Gamma_0$ or at $\Gamma_m$. Indeed, since $\see$ is assumed to be canonical, the non-degenerate members have end-points strictly between $\gamma_0$ and $\gamma_m$ and no degenerate member intersects $\Gamma_0$ or $\Gamma_m$.

For (T1), let $e_i\notin P$ and let $s$ be a member open at $\Gamma_{i-1}$ but not at $\Gamma_i$, with endpoints $u$ and $v$. Then $\gamma_{i-1}<x(v)\le \gamma_i$, and $v$ lies in $P$, so $v$ is an event with $x$-coordinate in that range and therefore $v=e_i$, contradicting $e_i\notin P$. The symmetric argument applies to a member open at $\Gamma_i$ but not at $\Gamma_{i-1}$, so $S_i=S_{i-1}$. Suppose now that two distinct lines of $R_i$ lie in $S_{i-1}$. Each carries a member open at both $\Gamma_{i-1}$ and $\Gamma_i$. Such a member contains every point of its line whose $x$-coordinate lies between those of its endpoints, in particular $e_i$. Two members would then intersect. Hence $|S_{i-1}\cap R_i|\le 1$.

For (T2), let $e_i\in P$ with owner $\ell$. A line $\ell'\notin R_i$ carries members whose endpoints lie in $P\cap\ell'$, hence differ from $e_i$ and, by the distinctness of $x$-coordinates, have $x$-coordinate different from $x(e_i)$; so no member on $\ell'$ has an endpoint with $x$-coordinate in $[\gamma_{i-1},\gamma_i]$, and $\ell'\in S_{i-1}$ if and only if $\ell'\in S_i$. Next let $\ell'\in S_{i-1}\cap R_i$ and let $s$ be the member on $\ell'$ open at $\Gamma_{i-1}$, with endpoints $u$ and $v$. Then $x(u)<\gamma_{i-1}<x(e_i)$, while $v\in P$ gives $x(v)>\gamma_{i-1}$ and hence $x(v)\ge x(e_i)$, because $e_i$ is the first event to the right of $\Gamma_{i-1}$. So $e_i$ lies on $\ell'$ with $x$-coordinate between those of $u$ and $v$, whence $e_i\in s$ and $\ell'=\ell$. The same argument at $\Gamma_i$ gives $S_i\cap R_i\subseteq\{\ell\}$.

For the cost, let $e_i\in P$, let $s$ be the member containing $e_i$, and let $u$ be the point of $P\cap s$ with the least $x$-coordinate. Since $\see$ is canonical, either $P\cap s=\{u\}$, and then $s$ is open at no separator, or $u$ is the endpoint of $s$ of least $x$-coordinate. In both cases $s$ is open at $\Gamma_{i-1}$ if and only if $u\neq e_i$, because the other endpoint of $s$ lies at least as far right as $e_i$, while $u$ is an event and $e_i$ is the first event to the right of $\Gamma_{i-1}$. The line of $s$ is $\mathtt{ow}(i)$, so $s$ is open at $\Gamma_{i-1}$ exactly when $\mathtt{ow}(i)\in S_{i-1}$; one direction is immediate and the other is the argument just given for (T2). Hence index $i$ is counted exactly when $e_i=u$, which happens once for each member of $\see$, and $c(\tau)=|\see|$.

Hence, we are done with proving that for every canonical solution, there is a trace of the same cost. Now, we show that for every trace, there is a solution of the same size.
\begin{clm}
    Every trace yields a solution of the same size.
\end{clm}
\begin{claimproof}
    Let $\tau$ be a trace. Call $i$ a \emph{start} when $e_i\in P$ and $\mathtt{ow}(i)\notin S_{i-1}$, and a \emph{continuation} when $e_i\in P$ and $\mathtt{ow}(i)\in S_{i-1}$. Every input-point index is of exactly one kind, and $c(\tau)$ is the number of starts. Fix a start $i$ and put $\ell:=\mathtt{ow}(i)$. Let $j(i)\ge i$ be the least index with $\ell\notin S_{j(i)}$; it exists because $S_m=\emptyset$, and $\ell\in S_r$ for every $r$ with $i\le r<j(i)$. Now, we consider two cases:

    \smallskip 

    \noindent \textbf{Case I ($j(i)>i$) :} Let $j=j(i)$. Then $\ell\in S_{j-1}\setminus S_j$, so $S_j\neq S_{j-1}$ and (T1) forces $e_j\in P$. Rule (T2) at $j$ gives $S_j\setminus R_j=S_{j-1}\setminus R_j$, which implies $\ell\in R_j$, and then $S_{j-1}\cap R_j\subseteq\{\mathtt{ow}(j)\}$ gives $\mathtt{ow}(j)=\ell$. In particular $j$ is a continuation, so no start equals $j$. Let $s_i:=\operatorname{conv}\{e_i,e_j\}$, a segment on $\ell$ with both distinct endpoints in $P$, and call it \emph{long}. Let $i<i'$ be starts whose members are long and lie on one line $\ell$, and write $j=j(i)$, $j'=j(i')$. If $j\ge i'$ then $i\le i'-1<j$, so $\ell\in S_{i'-1}$, contradicting that $i'$ is a start; hence $j<i'$ and the two members are non-intersecting. Now, let $e_i$ lie on at least two lines of $\lee$ and let $s_{i_1}$ be a long member containing $e_i$, on the line $\ell_1\in R_i$ and with $x(e_{i_1})\le x(e_i)\le x(e_{j_1})$, where $j_1=j(i_1)$. If $e_i\notin P$, then $e_i$ is neither $e_{i_1}$ nor $e_{j_1}$, both of which lie in $P$, so $i_1<i<j_1$ and $\ell_1\in S_{i-1}\cap R_i$; rule (T1) gives $|S_{i-1}\cap R_i|\le 1$, so $\ell_1$ does not depend on the choice of $s_{i_1}$. If $e_i\in P$, then either $i_1=i$ or $j_1=i$, and $\ell_1=\mathtt{ow}(i)$ by the construction, or $i_1<i<j_1$ and $\ell_1\in S_{i-1}\cap R_i\subseteq\{\mathtt{ow}(i)\}$ by (T2); again $\ell_1$ does not depend on the choice. In both cases, all long members containing $e_i$ lie on one line, and at most one of them contains $e_i$. So the long members are pairwise non-intersecting.

   \smallskip 

    \noindent \textbf{Case II ($j(i)=i$) :} Then $e_i$ lies on no long member: if $s_{i_1}$ were one, then its line $\ell_1$ lies in $R_i$, and $i_1\ne i$ because $j(i_1)>i_1$ while $j(i)=i$, and $j_1\ne i$ because $j_1$ is a continuation; so $i_1<i<j_1$, giving $\ell_1\in S_{i-1}\cap R_i$ and, by (T2), $\ell_1=\mathtt{ow}(i)$, so that $\mathtt{ow}(i)\in S_{i-1}$, contradicting that $i$ is a start. Now choose one $\delta>0$ smaller than the distance from $e_i$ to the union of the long members, than the distance from $e_i$ to any event other than $e_i$, and than half the distance between $e_i$ and $e_{i'}$ for any two such indices $i\neq i'$. For each start $i$ with $j(i)=i$ let $s_i$ be a segment on $\mathtt{ow}(i)$ that contains $e_i$ and has diameter less than $\delta$; call it \emph{short}. By the choice of $\delta$ a short member meets no long member and no other short member, and the only event it contains is $e_i$.

    Let $\mathcal F$ consist of the long and the short members. It is pairwise non-intersecting, by the two paragraphs above. Distinct starts give distinct members, since members coming from distinct starts are non-intersecting and nonempty, so $|\mathcal F|=c(\tau)$.

    Finally $\mathcal F$ contains $P$. Let $p=e_i\in P$ with owner $\ell$. If $i$ is a start, then $p$ is an endpoint of $s_i$. Otherwise $\ell\in S_{i-1}$, so some start $i'<i$ has $\ell\in S_r$ for all $i'\le r<j'$ with $j'\ge i$, and then $x(e_{i'})<x(p)\le x(e_{j'})$ places $p$ on $\ell$ between the endpoints of $s_{i'}$, so $p\in s_{i'}$.
    \hfill$\diamond$
\end{claimproof}
Hence we have $\min_\tau c(\tau)=\OPT(P,\lee)$.\lineqed
\end{proof}

\subsubsection{The Dynamic Program.}
For $i\in\{0,\dots,m\}$ and $S\subseteq\lee$, let $\DP[i,S]$ be the minimum cost of a prefix $(S_0,\dots,S_i;\mathtt{ow})$ that satisfies $S_0=\emptyset$, $S_i=S$, and rules (T1) and (T2) at every index up to $i$, the cost of a prefix is $\bigl|\{r\le i\,|\,e_r\in P \text{ and } \mathtt{ow}(r)\notin S_{r-1}\}\bigr|$. Set $\DP[i,S]=\infty$ when no such prefix exists. We define, $\DP[0,\emptyset]=0$ and $\DP[0,S]=\infty$ for $S\neq\emptyset$.

At an ordinary event, rule (T1) forbids a change of the set and allows at most one line of $R_i$ to stay open, so
\[
    \DP[i,S] \;=\;
    \begin{cases}
        \DP[i-1,S], & \text{if } |S\cap R_i|\le 1,\\[1mm]
        \infty,     & \text{otherwise.}
    \end{cases}
\]

At an input-point event, rule (T2) leaves the lines outside $R_i$ untouched and allows at most the owner inside $R_i$, on either side. Given the new set $S'$ and a candidate owner $\ell\in R_i$, the old set is therefore $S'\setminus R_i$ or $(S'\setminus R_i)\cup\{\ell\}$, and no other choice is admissible. Hence
\[
    \DP[i,S'] \;=\;
    \min_{\substack{\ell\in R_i\\ S'\cap R_i\subseteq\{\ell\}}}\;
    \min_{S\in\{\,S'\setminus R_i,\;(S'\setminus R_i)\cup\{\ell\}\,\}}
    \Bigl(\DP[i-1,S]+\mathbf{1}[\ell\notin S]\Bigr).
\]
The increment is $\mathbf{1}$ exactly when no member on the owner arrives from the left, that is, when a new member starts at $e_i$, and it is $\mathbf{0}$ when a member on the owner is already open and either stops at $e_i$ or continues past it.

\begin{lemma}
\label{thm:sweep-dp}
The dynamic program returns $\DP[m,\emptyset]=\OPT(P,\lee)$, and an optimal family of segments can be recovered from it.
\end{lemma}

\begin{proof}
The two recurrences enumerate, for each index, exactly the transitions allowed by (T1) and (T2), and they add the contribution of the index to $c(\tau)$. So $\DP[i,S]$ is the least cost of a prefix ending in $S$, by induction on $i$, and $\DP[m,\emptyset]$ is the least cost of a trace. \Cref{thm:trace} gives $\DP[m,\emptyset]=\OPT(P,\lee)$.

For the family itself, store with each successful relaxation a pointer to the entry it came from, together with the owner used. Backtracking from $(m,\emptyset)$ produces a trace of that cost, and the second half of the proof of \Cref{thm:trace} turns it into a solution of the same size.\lineqed
\end{proof}

\noindent \textbf{Time Complexity.} There are $m+1=\bigO(n+t^2)$ separators and at most $2^t$ states at each, so the table has $\bigO\bigl((n+t^2)2^t\bigr)$ entries. An ordinary event needs the size of $S\cap R_i$, which costs $\bigO(t)$. An input-point event tries at most $|R_i|\le t$ owners and, for each, two predecessor sets, which also costs $\bigO(t)$. Hence total running time is 
$
    \bigO\bigl((n+t^2)2^t\bigr)\cdot\bigO(t)
    \;=\;
    \bigO\bigl((n+t^2)\,t\,2^t\bigr).
$
Thus, we get the following theorem.

\begin{theo}
\label{thm:lr-t-algorithm}
\LRSC{} can be solved in $2^{\bigO(t)}n^{\bigO(1)}$ time.
\end{theo}

\subsection{FPT algorithm parameterized by $k$} \label{sec:fpt2}

\begin{theo}
\label{thm:lr-k-algorithm}
\LRSC~can be solved in
$k^{\bigO(k)}n^{\bigO(1)}$ time.
\end{theo}
\begin{proof}
    We use the same algorithm as in \Cref{thm:main}, with one additional step. Let $\aee$ denote the algorithm from \Cref{thm:main}. For every successful branch of $\aee$, we check whether every segment produced by that branch is supported by a line of $\lee$. If so, we accept; otherwise, we discard that branch. This additional line-membership check takes $t^{\bigO(1)}$ time. Since, without loss of generality, we may assume that $t=\bigO(n^2)$, the overall running time remains $k^{\bigO(k)}n^{\bigO(1)}.$

    For correctness, observe that if $(P,k,\lee)$ is a \yes-instance of \LRSC, then $(P,k)$ is also a \yes-instance of \SC. Thus, a solution of $(P,k,\lee)$ can be obtained by generating the corresponding solution of $(P,k)$ with the additional line-membership checking.
\end{proof}

\section{Relation to Point Line Cover}\label{sec:linecover}

Let $\lc(P)$ be the minimum number of lines whose union contains $P$.  The supporting lines of any segment cover form a line cover, after choosing an arbitrary supporting line for each degenerate segment.  Hence $\lc(P)\le\sco(P)$.

\begin{lemma}\label{lem:insertion}
If the lines $\ell_1,\ldots,\ell_r$ cover $P$, then $P$ has a segment cover of size at most $r(r+1)/2$. Such a cover can be constructed in polynomial time.
\end{lemma}
\begin{proof}
Assign each point to the first line in the ordering that contains it. On $\ell_i$, the previous $i-1$ lines split it into at most $i$ open intervals. An assigned point cannot lie at an intersection with an earlier line, since such a point would already have been assigned to that earlier line. For every interval containing assigned points, take the segment between the leftmost and rightmost assigned points. If the interval contains only one assigned point, use that point as a degenerate segment.

The segments on $\ell_i$ do not intersect any earlier line. Therefore, all the segments are pairwise non-intersecting. They cover all the assigned points and use at most $\sum_{i=1}^r i=\frac{r(r+1)}{2}$ segments.\lineqed
\end{proof}

\begin{corollary}\label{cor:sandwich}
For every finite point set $P$, $\lc(P)\le\sco(P)\le\frac{\lc(P)(\lc(P)+1)}2$. Moreover, the upper bound is tight: for every integer $r\geq 1$, there exists
a finite point set $P$ such that  $\lc(P) =r$ and $\sco(P) = r(r + 1)/2$.
\end{corollary}
\begin{proof}
The first part follows trivially from \Cref{lem:insertion}.

Take $r$ lines $\ell_1,\ldots,\ell_r$ such that no two are parallel and no three meet at the same point. The other $r-1$ lines divide each $\ell_i$ into $r$ open pieces. Put $ M=\frac{r(r+1)}2 $ points in each piece, avoiding all intersection points. A line different from the $\ell_i$ meets $\ell_i$ in at most one point. Hence, $r-1$ lines can cover at most $r-1$ points of any $\ell_i$, while $\ell_i$ contains $rM$ points. Thus, $r-1$ lines cannot cover all points of $P$. Therefore, $\lambda(P)\ge r$. On the other hand, the $r$ lines $\ell_1,\ldots,\ell_r$ cover all points of $P$, so $\lambda(P)=r$.

Now let $\see$ be a pairwise non-intersecting cover of $P$ with $k<r(r+1)/2$ members. Fix a line $\ell_i$. A member of $\see$ that is not contained in $\ell_i$ can cover at most one point of $\ell_i$. Since every piece of $\ell_i$ contains $M>k$ points, each piece must contain a point covered by a member of $\see$ that is contained in $\ell_i$. Let $b_i$ be the number of intersection points on $\ell_i$ that are not contained in any member of $\see$ lying on $\ell_i$. At each such intersection point, the members of $\see$ contained in $\ell_i$ must stop and start again: otherwise, one member would pass through the intersection point. Hence, $\ell_i$ contains at least $1+b_i$ members of $\see$. Consider an intersection point of $\ell_i$ and $\ell_j$. Since the members of $\see$ are pairwise non-intersecting, they cannot both contain this intersection point. Thus, this intersection point is counted in $b_i$ or in $b_j$. Therefore, $ \sum_{i=1}^r b_i\ge \binom r2. $ Consequently, $ |\see| \ge \sum_{i=1}^r(1+b_i) = r+\sum_{i=1}^r b_i \ge r+\binom r2 =\frac{r(r+1)}2. $ This contradicts $k<r(r+1)/2$. Hence, $ \zeta(P)\ge \frac{r(r+1)}2. $ \Cref{lem:insertion} gives the reverse inequality, and therefore $ \zeta(P)=\frac{r(r+1)}2. $\lineqed
\end{proof}

% \begin{tcolorbox}[
%     enhanced,
%     title={\color{black}\textsc{Point Line Cover}},
%     colback=white,
%     boxrule=0.4pt,
%     attach boxed title to top center={
%         xshift=-3.5cm,yshift*=-2.5mm
%     },
%     boxed title style={
%         size=small,
%         frame hidden,
%         colback=white
%     }
% ]
% \vspace{-1mm}
% \textbf{Input:} A finite point set $P\subseteq\mathbb{R}^{2}$, and a non-negative integer $k$.

% \textbf{Question:} Can $P$ be covered by at most $k$ lines?
% \end{tcolorbox}

\begin{problemm}
  \problemtitle{\textsc{Point Line Cover}}
  \probleminput{A finite point set $P\subseteq\mathbb{R}^{2}$, and a non-negative integer $k$.}
  \problemquestion{Can $P$ be covered by at most $k$ lines?}
\end{problemm}

% \begin{tcolorbox}[masterbox,title={\sc Point Line Cover}]
% \begin{tabular}{@{}lp{10cm}@{}}
% \textit{Input:}
% & A finite point set $P\subseteq\mathbb{R}^{2}$, and a non-negative integer $k$.\\[3pt]

% \textit{Question:}
% & Can $P$ be covered by at most $k$ lines?
% \end{tabular}
% \end{tcolorbox}

\noindent The following result is known due to Langerman and Morin~\cite{DBLP:journals/dcg/LangermanM05}.

\begin{proposition}[\cite{DBLP:journals/dcg/LangermanM05}]\label{lem:plc}
    {\sc Point Line Cover} is solvable in $k^{\bigO(k)}+n^{\bigO(1)}$ time.    
\end{proposition}

\begin{theo}\label{thm:lambda-fpt}
$\SC$ can be solved in time $\lc(P)^{\bigO(\lc(P)^2)}n^{\bigO(1)}$.
\end{theo}
\begin{proof}
Assume $P\ne\emptyset$ and so $r:=\lc(P)\ge1$. Compute $r$ and a minimum line cover using \Cref{lem:plc}. By \Cref{cor:sandwich}, $r\le \sco(P)\le U:=\frac{r(r+1)}2$.
For a decision threshold $k<r$, reject.  For $k\ge U$, accept and construct a cover using \Cref{lem:insertion}.  In the only remaining case, $r\le k<U=\bigO(r^2)$, applying \Cref{thm:main};  running time becomes $k^{\bigO(k)}n^{\bigO(1)} =r^{\bigO(r^2)}n^{\bigO(1)}=\lc(P)^{\bigO(\lc(P)^2)}n^{\bigO(1)}$.\lineqed
\end{proof}

\section{Conclusion}\label{sec:conclusion}
We obtained improved FPT algorithms for \SC{} and \LRSC{}, an $\bigO(t^2)$-vertex kernel for \LRSC{}, and an FPT algorithm for \SC{} parameterized by $\lc(P)$. Several questions remain open, including single-exponential algorithms and polynomial kernels for \SC{}.

\bibliographystyle{plainurl}
\bibliography{00ref} 
% \appendix
% \input{appendix}
\end{document}